\documentclass[aps,pre,preprint,superscriptaddress,nofootinbib]{revtex4-1}
\usepackage[T1]{fontenc}
\usepackage[utf8]{inputenc}
\usepackage{color}
\usepackage{float}
\usepackage{amsmath}
\usepackage{amssymb}
\usepackage{graphicx}
\usepackage{hyperref}
\usepackage{appendix}
\usepackage{xcolor}
\newcommand{\jingma}[1]{{#1}}
\usepackage{comment}

\makeatletter

\@ifundefined{showcaptionsetup}{}{
 \PassOptionsToPackage{caption=false}{subfig}}
\usepackage{subfig}
\makeatother

\begin{document}
\renewcommand{\labelenumii}{\arabic{enumi}.\arabic{enumii}}

\title{Peak fraction of infected in epidemic spreading for multi-community networks}

\author{Jing Ma}
\email{jingma@bu.edu}
\affiliation{Department of Physics, Boston University, Boston, MA 02215, USA}

\author{Xiangyi Meng}
\affiliation{Center for Complex Network Research and Department of Physics, Northeastern University, Boston, MA 02115, USA}

\author{Lidia A. Braunstein}
\affiliation{Instituto de Investigaciones
  F\'isicas de Mar del Plata (IFIMAR), FCEyN,
  Universidad Nacional de Mar del Plata-CONICET, D\'ean Funes 3350,
  (7600) Mar del Plata, Argentina}
\affiliation{Department of Physics, Boston University, Boston, MA 02215, USA}

\begin{abstract}

One of the most effective strategies to mitigate the global spreading of a pandemic (e.g., COVID-19) is to shut down international airports.
From a network theory perspective,
this is since international airports and flights,
essentially playing the roles of bridge nodes and bridge links between countries as individual communities,
dominate the epidemic spreading characteristics in the whole multi-community system.
Among all epidemic characteristics,
the peak fraction of infected,
$I_{\max}$,
is a decisive factor in evaluating an epidemic strategy given limited capacity of medical resources,
but is seldom considered in multi-community models.
In this paper,
we study a general two-community system interconnected by a fraction $r$ of bridge nodes and its dynamic properties,
especially $I_{\max}$,
under the evolution of the Susceptible-Infected-Recovered (SIR) model.
Comparing the characteristic time scales of different parts of the system allows us to analytically derive the asymptotic behavior of $I_{\max}$ with $r$,
as $r\rightarrow 0$,
which follows different power-law relations in each regime of the phase diagram.
We also detect crossovers when $I_{\max}$ changes from one power law to another,
crossing different power-law regimes as driven by $r$.
Our results enable a better prediction of the effectiveness of strategies acting on bridge nodes,
denoted by the power-law exponent $\epsilon_I$ as in $I_{\max}\propto r^{1/\epsilon_I}$.  

\end{abstract}

\maketitle


\section{Introduction}

Network science has provided many useful tools for studying epidemic problems
\cite{newman2002spread}.
By modeling an epidemic-confronting society as a network,
where each individual is modeled as a node and \jingma{all physical contacts between individuals so that the disease might get transmitted as links,}
an epidemic problem can often be reduced to a pure problem of percolation theory and network dynamics which strongly depend on the network topology.
In many synthetic and real-world complex networks,
it is known that the number of short loops is negligible \cite{cohen2011resilience},
and thus the network topology can be characterized by two generating functions $G_0$ and $G_1$ denoting the \emph{degree distribution} and the \emph{excess degree distribution}:
$G_0(x)=\sum_k P(k) x^k$ and $G_1(x)=\langle k\rangle^{-1}\sum_k kP(k)x^{k-1}$,
respectively,
given $P(k)$ the fraction of nodes of degree $k$ in the network,
\jingma{and $\langle k \rangle = \sum_k kP(k)$ the average degree}
\cite{newman2001random,callaway2000network,dorogovtsev2002evolution}.

In the Susceptible-Infected-Recovered (SIR) model,
the course of a disease can be modeled as three states,
and each individual can be in one of these three states at any instant:
susceptible (S, i.e., not infected yet),
infected (I),
and recovered (R) \cite{istvan2017mathematics}.
An individual will recover $t_r$ time steps after being infected, and is then immune to the disease and will never get infected again.
\jingma{Note} that the final steady state of the SIR model can be mapped into a link percolation problem
\cite{grassberger1983critical,newman2010networks,stauffer2018introduction,pastor2015epidemic,mello2021epidemics,sander2002percolation}.
In this mapping,
the fraction of individuals that have ever been infected at the final state $R_\text{final}$ is just the size of the cluster that patient zero belongs to in the link percolation problem,
which is the order parameter of a phase transition;
the transmissibility $T$ in the SIR model, which is the probability that an infected node can spread this disease to its neighbor through a link before it recovers, is equivalent to the probability of a link being occupied in the link percolation problem,
which is the control parameter.

\jingma{In recent years,
there are many studies about epidemics in systems with more complicated structures,
such as multi-group modelling or multi-community systems.
In the multi-group modelling,
nodes are classified into different groups based on age or other factors \cite{bajiya2021global,feng2005global}.
In systems of multiple communities,
each community is itself a complex network of some degree distribution,
while multiple communities are coupled to each other through either shared nodes
\cite{buldyrev2010catastrophic,son2012percolation,kryven2019bond},
or bridge links that follow a possibly different degree distribution \cite{gao2011robustness,kenett2014network,gao2013percolation,dong2018resilience}.}
\jingma{
The structure of multi-community systems with bridge links is used in our research since it better reflects the real world.
In practice,
different countries (represented by communities) may have different transportation capabilities,
and thus their own topological properties.
Also,
constraints on international traveling are usually more strict than domestic ones,
so it is necessary to distinguish transmissibilities along bridge links between communities from internal links \cite{pham2021estimating,adekunle2020delaying}.}
In a system of multiple communities connected by bridge links,
which allows for different transmissibilities along internal links ($T^i$) and bridge links ($T^b$),
it has been shown that $R_\text{final}$ asymptotically follows different power-law behaviors with $r$ in different regimes,
where $r$ is the fraction of nodes in the whole system that are bridge nodes (nodes with bridge links attached)
\cite{ma2020role}.
These results enable better decisions about epidemic strategies
such as whether social distancing strategies are needed (to reduce transmissibility $T$) or how many international airports need to be closed (to reduce the fraction of bridge nodes $r$).

Besides the final steady state $R_\text{final}$,
the dynamic properties of the SIR process,
especially the peak \jingma{fraction of infected} $I_{\max}$,
are also of great interest.
The dynamics of SIR has been well known to belong to the same dynamic universality class of link percolation,
given its equivalence to the breadth-first process (the Leath-Alexandrowicz algorithm \cite{leath1976cluster,alexandrowicz1980critically}) that is used for simulating the growth of percolation clusters \cite{zhou2012shortest}.
In this paper,
instead of looking at the final state of SIR,
we study its dynamic properties in a two-community system with bridge links.
By comparing the time scale of different parts of the system,
we find that the peak \jingma{fraction of infected} $I_{\max}$ also follows different power laws with the fraction of bridge nodes $r$ in different regimes as $r\rightarrow 0$.
The regimes are determined by the comparison between the order parameters ($T^i$ and $T^b$) and their critical values in isolated systems,
while the exponents in different regimes are related to the exponents for $R_\text{final}$ \jingma{\cite{ma2020role}}.
All of our results are verified by numerical simulations.
\jingma{Now, we can predict not only the total number of individuals ever been infected in the SIR model \cite{ma2020role}},
but also the maximum number of \jingma{infected} during the epidemic.
In practice,
$I_{\max}$ is \jingma{the} more decisive factor,
as it actually decides the transient maximum capacity of patients who can receive timely treatment.


\section{Model}

\begin{figure}[h]
\centering
\includegraphics[width=0.3\textwidth]{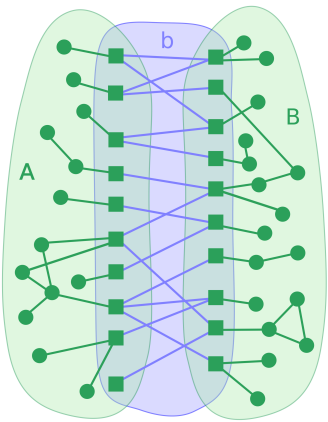}
\caption{Illustration of a two-community system composed by communities $A$ and $B$. The subsystem $b$ is composed of bridge nodes and bridge links. Bridge nodes are denoted by squares and internal nodes by circles.}
\label{fig_topology}
\end{figure}

Consider a system of two communities $A$ and $B$,
where a fraction $r$ of nodes from each community are \emph{bridge nodes},
between which \emph{bridge links} that interconnect $A$ and $B$ exist.
The subsystem composed of bridge nodes and bridge links is denoted by $b$,
as shown in Fig.~\ref{fig_topology}.
\jingma{Both communities and the bridge links are generated by the configuration model and are guaranteed uncorrelated \cite{dorogovtsev2010lectures}.}
For simplicity,
we assume the two communities $A$ and $B$ are statistically identical,
so that $P^A(k)=P^B(k)\equiv P^i(k)$,
and that the internal transmissibility is also the same within each community,
given by $T^A(k)=T^B(k)\equiv T^i(k)$.
The bridge links are allowed to have a different degree distribution $P^b(k)$ and a different transmissibility $T^b$.
Note that all the methods and results in this paper can be generalized to cases with $P^A(k)\ne P^B(k)$ and/or $T^A\ne T^B$.

\begin{figure}[h]
\centering
\includegraphics[width=0.49\textwidth]{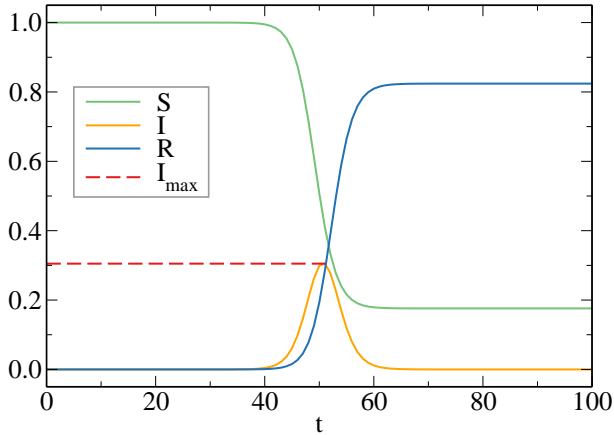}
\caption{Time dependence of the susceptible (S), infected (I), and recovered (R) in the SIR model in a two-community system, where both internal and bridge links follow Poisson distributions, with $\langle k^i \rangle = 4$, $\langle k^b \rangle = 10$, $r=0.1$, $T^i=0.5$, $T^b=0.2$.}
\label{fig_sir}
\end{figure}

The step-by-step evolution of the system can then be simulated by the Edge-Based Compartmental Model (EBCM) adapted to the SIR model \cite{miller2012edge,valdez2018role}.
\jingma{
The EBCM is a set of difference equations that can reproduce the evolution of disease spreading,
i.e.,
the time dependence of the fraction of susceptible $S$,
the fraction of infected $I$,
and the fraction of recovered $R$,}
using much less time than calculating the states of all nodes individually at each time step
(see Appendix \ref{app_ebcm}).
For example,
in a two-community system 
where both internal and bridge links follow Poisson distributions $P(k)=\langle k \rangle^k e^{-\langle k \rangle}/k!$,
with $\langle k^i \rangle = 4$, $\langle k^b \rangle = 10$, $r=0.1$, $T^i=0.5$, $T^b=0.2$,
the time dependence of $S$, $I$, $R$ based on the EBCM simulation [Eqs.~(\ref{eq_difference_thetai})-(\ref{eq_difference_deltaib})] shows that $R$ will increase from zero and then stabilize to a value $R_\text{final}$,
and that $I$ will increase at the beginning but then decrease after passing a peak value $I_{\max}$ (Fig.~\ref{fig_sir}).
\jingma{It has been shown that $R_\text{final}$ has different power-law behaviors with $r$ in different regimes \cite{ma2020role}.}
In this paper,
we will show that $I_{\max}$ also follows power-law relations with $r$ as $r\rightarrow 0$,
and that crossovers exist between some regimes when $r$ is not small enough.

\section{Asymptotic dependence of $I_{\max}$ on $r$ in different regimes}

By mapping the SIR model to a link percolation problem,
we can apply well-known results of percolation theory to epidemic problems.
Hence,
we are going to use the terminologies in the SIR model and percolation theory interchangeably.
In the SIR model,
the critical value of transmissibility in an isolated network is given by $T_c=1/(\kappa-1)$,
where $\kappa=\langle k^2 \rangle/\langle k \rangle$ is the branching factor \cite{lagorio2011quarantine,buono2014epidemics}.
This critical point is characterized by many behaviors,
e.g., the probability to find a cluster of size $s$ is given by $P(s) \sim s^{-\tau+1}\exp(-s/s_{\max})$, where $s_{\max}\sim |T-T_c|^{-1/\sigma}$ is the largest finite cluster size \cite{newman2010networks,cohen2002percolation}.
For Erd\"os-R\'enyi (ER) networks whose degree distribution follows a Poisson distribution $P(k)=\langle k \rangle^k e^{-\langle k \rangle}/k!$,
we always have $\tau=5/2$;
for scale-free (SF) networks where the degree distribution is a power law $P(k)\propto k^{-\lambda}$ with $3<\lambda<4$,
$\tau$ is given by $\tau = (2\lambda-3)/(\lambda-2)$ \cite{cohen2003structural}.
Also,
the correlation length $\xi$ diverges around the critical point following $\xi\sim|T-T_c|^{-\nu}$,
where $\nu=1/2$ for both ER and SF networks.
There are also dynamic behaviors around the critical point,
e.g., the chemical distance $l$,
which represents the time scale in epidemic models \cite{grassberger1992spreading},
is related to the correlation length $\xi$ by $l\sim \xi^z \sim|T-T_c|^{-z\nu}$,
in which $z=2$ for both ER and SF networks.

Due to the abrupt change in behaviors around the critical points,
we are going to split the space of the combination of $T^i$ and $T^b$ into seven regimes \jingma{\cite{ma2020role}},
based on whether $T^i$ is less than, equal to, or larger than $1/(\kappa^i-1)$,
and whether $T^b$ is less than, equal to, or larger than $1/(\kappa^b-1)$
\jingma{(Fig.~\ref{fig_regimes})}.
Note that $1/(\kappa^i-1)$ or $1/(\kappa^b-1)$ is the critical value of $T^i$ or $T^b$ when the respective part is \jingma{isolated,
and} we are going to look at the peak \jingma{fraction of infected} $I_{\max}$ in each regime.

In order to derive the behavior of $I_{\max}$,
it is helpful to denote $I^b$ as the fraction of bridge nodes that are infected at any instant,
and $I^b_{\max}$ as the peak \jingma{fraction of infected} for bridge nodes.
For a community,
the peak \jingma{fraction of infected} $I_{\max}$ is related to the status of its bridge nodes,
i.e., either $rR^b_\text{final}$ or $rI^b_{\max}$,
where $R^b_\text{final}$ is the fraction of bridge nodes that are recovered at the final state,
depending on whether they get infected within a small or large time scale.
Specifically,
if the time scale of a community is much less than the time scale of the whole system,
the spreading of the disease in the community can be treated as multiple ``breakouts'' within the community occurring one after another,
i.e., those ``breakouts'' will not overlap over time;
on the other hand,
if the time scale of a community is much larger than the time scale of the system,
all the ``breakouts'' will keep spreading within the community and accumulate over time.

The dependencies of $I_{\max}$ and $I^b_{\max}$ in each regime are discussed separately as follows:
\begin{enumerate}
    \item
    When $T^i<1/(\kappa^i-1)$,
    \jingma{there are at most a fraction $rI_{\max}^b$ of bridge nodes that are infected at any instant,
    and each of the nodes is expected to expand to at most $s_{\max}$ nodes within the community,
    where $s_{\max}$ is the largest finite cluster size.}
    Consequently,
    $rI_{\max}^b \le I_{\max} \le rI_{\max}^b\cdot s_{\max}$,
    where $s_{\max}$ is finite,
    so $I_{\max} \propto rI_{\max}^b$,
    as $r \rightarrow 0$,
    which is true for any value of $T^b$.
    \begin{enumerate}
        \item{(Regime I)}
        When $T^b<1/(\kappa^b-1)$,
        the whole system is in non-epidemic regime,
        so $I_{\max}$ or $I^b_{\max}$ is not a power law of $r$.
        \item{(Regime II)}
        When $T^b=1/(\kappa^b-1)$,
        there is the relation $R^b_\text{final}\sim I^b_{\max}\cdot l^b$ \jingma{due to $R_\text{final}^b \sim \int I^b \cdot\text{d}l^b$},
        where $l^b$ represents the time scale of the bridge link part.
        Since $l^b\sim |T^b-T^b_c|^{-z^b\nu^b}\sim |T^b-T^b_c|^{-1}$ given $z=2$ and $\nu=1/2$ for both ER and SF networks with $3<\lambda<4$,
        and also $|T^b-T^b_c|\sim r$ where $T^b_c$ is the critical value of bridge link transmissibility for the whole system given a fixed value of $T^i$ (see Appendix \ref{app_sec_scale}),
        we have $I^b_{\max}\sim rR^b_\text{final}$.
        \item{(Regime III)}
        When $T^b>1/(\kappa^b-1)$,
        there is a giant component within the network of bridge links $b$ in finite time steps,
        so $I^b_{\max}$ is not a power law of $r$.

    \end{enumerate}
    \item{(Regime IV, V, VI)}
    When $T^i=1/(\kappa^i-1)$,
    \jingma{each community has a divergent time scale,
    while the whole system has a finite time scale
    \footnote[2]{\jingma{The time scale for a system is divergent at critical since $l\sim|T-T_c|^{-z\nu}$. However, the relation holds true only when the system size is large enough compared with the correlation length $\xi$. For percolation in an infinitely large system starting from a small fraction $I(0)$ of nodes, the time scale $l$ at critical depends on $I(0)$ and keeps increasing to infinity as $I(0)$ decreases; but $l$ is finite and independent of $I(0)$ as long as $I(0)$ is small enough, if above critical. For percolation in a finite system, we cannot really have infinite time scale even if we are at critical, since the system size will be considered not large enough and the relation $l\sim|T-T_c|^{-z\nu}$ no longer holds as $T$ goes too close to $T_c$. However, the time scale near critical will keep increasing as the system gets larger, while it will be independent of the system size when above critical, as long as the system is large enough. In either case, we can always say the time scale $l$ at critical is significantly larger than when it is above critical, for reasonably large systems.}}.
    }
    There are $rR^b_\text{final}$ bridge nodes being infected in total,
    and they get infected within a short period of time due to the finite time scale of the system.
    This is equivalent to considering that there are $rR^b_\text{final}$ bridge nodes that get infected at the same time,
    and they are going to spread the disease within each community,
    so it is expected that there are at most $I_{\max} \propto rR^b_\text{final}$ nodes that are being infected at the same time (see Appendix \ref{app_sec_multiple}).
    \item{(Regime VII)}
    When $T^i>1/(\kappa^i-1)$,
    there is always a giant component within each community,
    and thus always a finite $I_{\max}$,
    which cannot be a power law of $r$.
    
\end{enumerate}

\begin{table}[htb]
\centering
\begin{tabular}{c c c c c c c} 
\hline
\hline
&\hspace*{5ex}& $T^b<\frac{1}{\kappa^b-1}$ &\hspace*{5ex}& $T^b=\frac{1}{\kappa^b-1}$ &\hspace*{5ex}& $T^b>\frac{1}{\kappa^b-1}$ \\ [1ex]
\hline
$T^i<\frac{1}{\kappa^i-1}$ &&
\begin{tabular}{@{}c@{}} $ $\vspace{-1ex} \\
$ $ \end{tabular} &&
\begin{tabular}{@{}c@{}} $I_{\max}\propto rI^b_{\max}\textbf{}$\vspace{-1ex} \\
$I^b_{\max}\propto rR^b_\text{final}$ \end{tabular} &&
\begin{tabular}{@{}c@{}} $I_{\max}\propto rI^b_{\max}$\vspace{-1ex} \\ 
$I^b_{\max}\propto 1$ \end{tabular} \\ [0ex]
\hline
$T^i=\frac{1}{\kappa^i-1}$ &&
$I_{\max}\propto rR^b_\text{final}$\vspace{-1ex} &&
$I_{\max}\propto rR^b_\text{final}$\vspace{-1ex} &&
$I_{\max}\propto rR^b_\text{final}$\vspace{-1ex} \\ [4ex]
\hline
\hline
\end{tabular}
\caption{Dependence of $I_{\max}$ and $I^b_{\max}$ in different regimes, as $r \rightarrow 0$.}
\label{app_tb_regimes_new}
\end{table}

\begin{table}[htb]
\centering
\begin{tabular}{c c c c c c c} 
\hline
\hline
&\hspace*{5ex}& $T^b<\frac{1}{\kappa^b-1}$ &\hspace*{5ex}& $T^b=\frac{1}{\kappa^b-1}$ &\hspace*{5ex}& $T^b>\frac{1}{\kappa^b-1}$ \\ [1ex]
\hline
$T^i<\frac{1}{\kappa^i-1}$ &&
\begin{tabular}{@{}c@{}}$ $\vspace{-1ex} \\ $ $\vspace{-1ex}
\end{tabular} &&
\begin{tabular}{@{}c@{}}$R_\text{final}\propto rR^b_\text{final}$\vspace{-1ex} \\ $R^b_\text{final}\propto (R_\text{final})^{\tau^b-2}$\vspace{-1ex}
\end{tabular} &&
\begin{tabular}{@{}c@{}}$R_\text{final}\propto rR^b_\text{final}$\vspace{-1ex} \\ $R^b_\text{final}\propto 1$\vspace{-1ex}
\end{tabular} \\ [4ex]
\hline
$T^i=\frac{1}{\kappa^i-1}$ &&
\begin{tabular}{@{}c@{}}$R_\text{final}\propto (rR^b_\text{final})^{\tau^i-2}$\vspace{-1ex} \\ $R^b_\text{final}\propto R_\text{final}$\vspace{-1ex}
\end{tabular} &&
\begin{tabular}{@{}c@{}}$R_\text{final}\propto (rR^b_\text{final})^{\tau^i-2}$\vspace{-1ex} \\ $R^b_\text{final}\propto (R_\text{final})^{\tau^b-2}$\vspace{-1ex}
\end{tabular} &&
\begin{tabular}{@{}c@{}}$R_\text{final}\propto (rR^b_\text{final})^{\tau^i-2}$\vspace{-1ex} \\ $R^b_\text{final}\propto 1$\vspace{-1ex}
\end{tabular} \\ [4ex]
\hline
\hline
\end{tabular}
\caption{Dependence of $R_\text{final}$ and $R^b_\text{final}$ in different regimes, as $r \rightarrow 0$ \cite{ma2020role}.}
\label{app_tb_regimes_old}
\end{table}

All the scaling relations are summarized in Table~\ref{app_tb_regimes_new}.
Combined with previously known results (Table~\ref{app_tb_regimes_old}),
we can find the asymptotic dependence of $I_{\max}$ on $r$,
as shown in Table \ref{tb_imaxexponents},
that gives the power-law exponent $\epsilon_I$ as in $I_{\max}\propto r^{1/\epsilon_I}$ in different regimes,
as $r \rightarrow 0$.

\begin{table}[htb]
\centering
\begin{tabular}{c c c c c c c} 
\hline
\hline
  &\hspace*{5ex}& $T^b<\frac{1}{\kappa^b-1}$ &\hspace*{5ex}& $T^b=\frac{1}{\kappa^b-1}$ &\hspace*{5ex}& $T^b>\frac{1}{\kappa^b-1}$ \\ [1ex]
\hline
$T^i<\frac{1}{\kappa^i-1}$ && \begin{tabular}{@{}c@{}}$\varnothing$\vspace{-1ex} \\ (Regime I) \end{tabular} && \begin{tabular}{@{}c@{}}$\epsilon_I=\frac{1-(\tau^b-2)}{2-(\tau^b-2)}$\vspace{-1ex} \\ (Regime II) \end{tabular} && \begin{tabular}{@{}c@{}}$\epsilon_I=1$\vspace{-1ex} \\ (Regime III) \end{tabular}\\ [4ex]
\hline
$T^i=\frac{1}{\kappa^i-1}$ &&
\begin{tabular}{@{}c@{}}$\epsilon_I=1-(\tau^i-2)$\vspace{-1ex} \\ (Regime IV) \end{tabular} &&
\begin{tabular}{@{}c@{}}$\epsilon_I=1-(\tau^i-2)(\tau^b-2)$\vspace{-1ex} \\ (Regime V) \end{tabular} && 
\begin{tabular}{@{}c@{}}$\epsilon_I=1$\vspace{-1ex} \\ (Regime VI) \end{tabular} \\ [4ex]
\hline
$T^i>\frac{1}{\kappa^i-1}$ && \begin{tabular}{@{}c@{}}$\varnothing$\vspace{-1ex} \\ (Regime VII) \end{tabular} && \begin{tabular}{@{}c@{}}$\varnothing$\vspace{-1ex} \\ (Regime VII) \end{tabular} && \begin{tabular}{@{}c@{}}$\varnothing$\vspace{-1ex} \\ (Regime VII) \end{tabular} \\ [1ex]
\hline
\hline
\end{tabular}
\caption{Power-law exponent $\epsilon_I$ as in $I_{\max}\propto r^{1/\epsilon_I}$ in different regimes, as $r \rightarrow 0$. $\varnothing$ denotes that there is no power-law relation in that regime.}
\label{tb_imaxexponents}
\end{table}

\begin{figure}[htb]
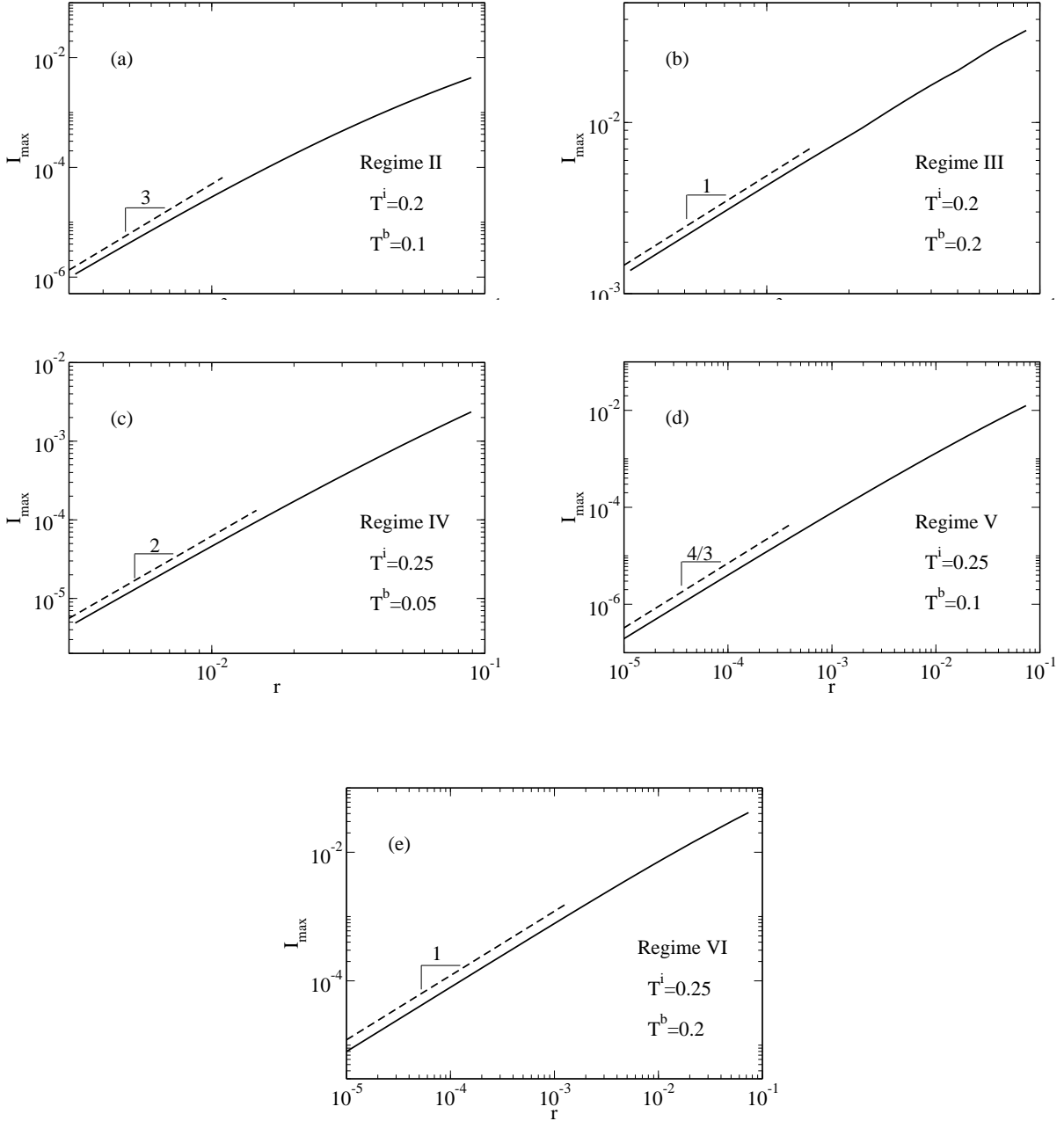

\begin{minipage}[t]{0.49\linewidth}\centering
\includegraphics[width=7.5cm]{{./figures/i_max_r_loglog_T_i_0.2_T_b_0.1}.eps}
\medskip
\end{minipage}\hfill
\begin{minipage}[t]{0.49\linewidth}\centering
\includegraphics[width=7.5cm]{{./figures/i_max_r_loglog_T_i_0.2_T_b_0.2}.eps}
\medskip
\end{minipage}\hfill
\begin{minipage}[t]{0.49\linewidth}\centering
\includegraphics[width=7.5cm]{{./figures/i_max_r_loglog_T_i_0.25_T_b_0.05}.eps}
\medskip
\end{minipage}\hfill
\vspace{1.5\baselineskip}
\begin{minipage}[t]{0.49\linewidth}\centering
\includegraphics[width=7.5cm]{{./figures/i_max_r_loglog_T_i_0.25_T_b_0.1}.eps}
\medskip
\end{minipage}\hfill
\begin{minipage}[t]{0.49\linewidth}\centering
\includegraphics[width=7.5cm]{{./figures/i_max_r_loglog_T_i_0.25_T_b_0.2}.eps}
\medskip
\end{minipage}
\caption{$I_{\max}$ as a function of $r$ for different regimes in an ER-ER system where a power law exists: (a) Regime II: $T^i=0.2$, $T^b=0.1$; (b) Regime III: $T^i=0.2$, $T^b=0.2$; (c) Regime IV: $T^i=0.25$, $T^b=0.05$; (d) Regime V: $T^i=0.25$, $T^b=0.1$; and (e) Regime VI: $T^i=0.25$, $T^b=0.2$. Both internal links and bridge links are ER networks, with $\langle k^i \rangle=4$ and $\langle k^b \rangle=10$,
such that $1/(\kappa^i-1)=0.25$ and $1/(\kappa^b-1)=0.1$,
respectively. In each regime, numerical solutions of Eqs.~(\ref{eq_difference_thetai})-(\ref{eq_difference_deltaib}) are plotted in solid lines, and dashed lines represent slopes predicted by Table~\ref{tb_imaxexponents}.}
\label{fig_exponents}
\end{figure}

\begin{figure}[htb]
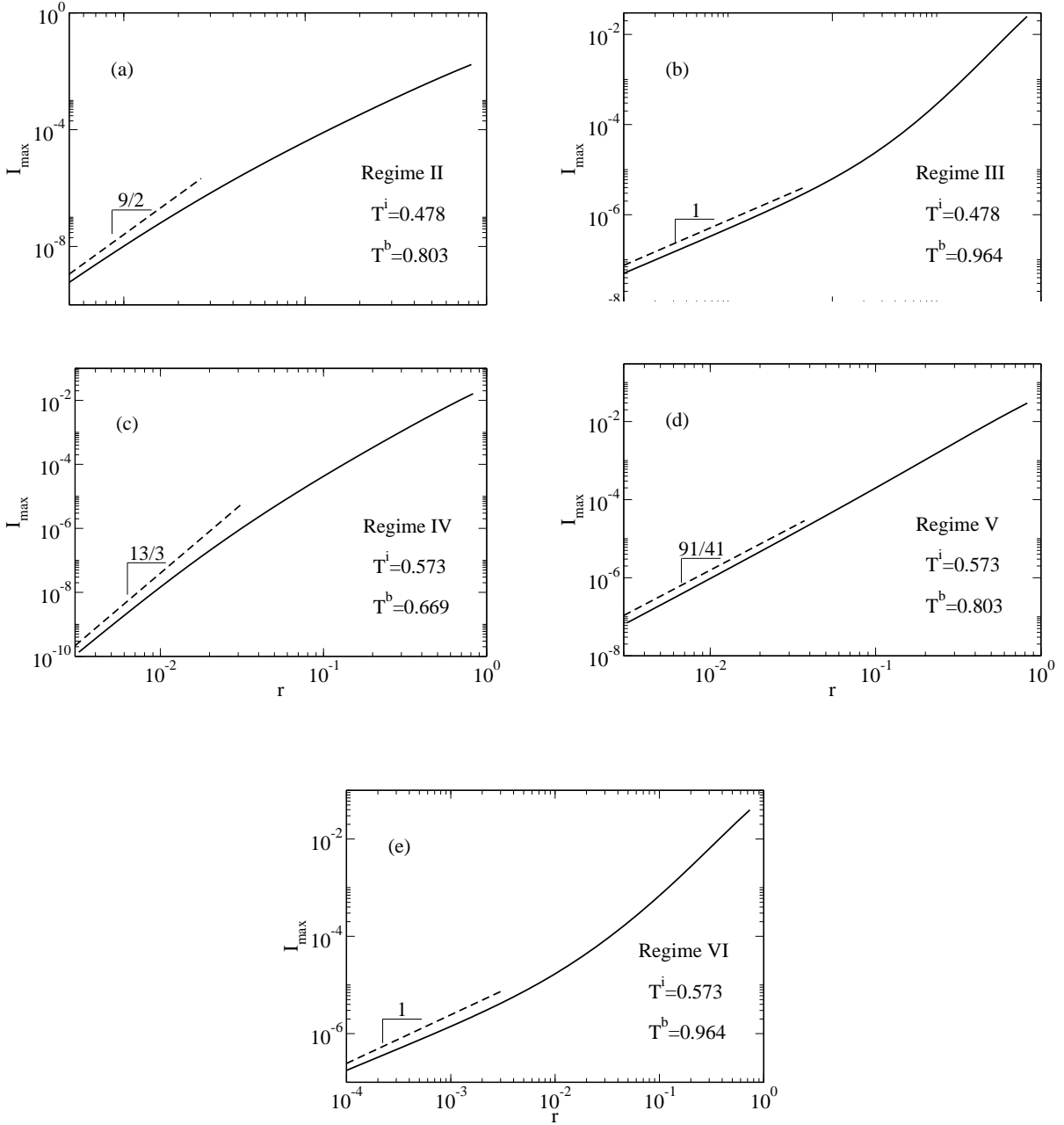

\begin{minipage}[t]{0.49\linewidth}\centering
\includegraphics[width=7.5cm]{{./figures/i_max_r_loglog_T_i_0.478_T_b_0.803}.eps}
\medskip
\end{minipage}\hfill
\begin{minipage}[t]{0.49\linewidth}\centering
\includegraphics[width=7.5cm]{{./figures/i_max_r_loglog_T_i_0.478_T_b_0.964}.eps}
\medskip
\end{minipage}\hfill
\begin{minipage}[t]{0.49\linewidth}\centering
\includegraphics[width=7.5cm]{{./figures/i_max_r_loglog_T_i_0.573_T_b_0.669}.eps}
\medskip
\end{minipage}\hfill
\vspace{1.5\baselineskip}
\begin{minipage}[t]{0.49\linewidth}\centering
\includegraphics[width=7.5cm]{{./figures/i_max_r_loglog_T_i_0.573_T_b_0.803}.eps}
\medskip
\end{minipage}\hfill
\begin{minipage}[t]{0.49\linewidth}\centering
\includegraphics[width=7.5cm]{{./figures/i_max_r_loglog_T_i_0.573_T_b_0.964}.eps}
\medskip
\end{minipage}
\caption{$I_{\max}$ as a function of $r$ for different regimes in a SF-SF system where a power law exists: (a) Regime II: $T^i=0.478$, $T^b=0.803$; (b) Regime III: $T^i=0.478$, $T^b=0.964$; (c) Regime IV: $T^i=0.573$, $T^b=0.669$; (d) Regime V: $T^i=0.573$, $T^b=0.803$; and (e) Regime VI: $T^i=0.573$, $T^b=0.964$. Both internal links and bridge links are SF networks, with $\lambda^i=3.3$ and $\lambda^b=3.4$,
such that $\tau^i=36/13$, $\tau^b=19/7$,
$1/(\kappa^i-1)=0.573$, and $1/(\kappa^b-1)=0.803$,
respectively. In each regime, numerical solutions of Eqs.~(\ref{eq_difference_thetai})-(\ref{eq_difference_deltaib}) are plotted in solid lines, and dashed lines represent slopes predicted by Table~\ref{tb_imaxexponents}.}
\label{fig_exponentssf}
\end{figure}

The results can be verified by comparing with the numerical solutions from the EBCM.
As in Fig.~\ref{fig_exponents},
in a system where both communities and bridge links are ER networks (ER-ER system) such that $\tau^i=\tau^b=5/2$,
numerical solutions of Eqs.~(\ref{eq_difference_thetai})-(\ref{eq_difference_deltaib}) are plotted in solid lines,
and the dashed lines are straight lines whose slopes are given by Table~\ref{tb_imaxexponents}.
It is clear that the numerical solutions agree with our prediction,
as $r\rightarrow 0$.
Our results also apply to SF networks with $3<\lambda<4$.
As in Fig.~\ref{fig_exponentssf},
in a system where both communities and bridge links are SF networks (SF-SF system)
with $\lambda^i=3.3$ and $\lambda^b=3.4$,
such that $\tau^i=36/13$ and $\tau^b=19/7$,
numerical solutions of Eqs.~(\ref{eq_difference_thetai})-(\ref{eq_difference_deltaib}) are plotted in solid lines,
and the dashed lines are straight lines predicted by Table~\ref{tb_imaxexponents}.
It is clear that the numerical solutions also agree with our prediction for SF networks as $r\rightarrow 0$.

Similar to our previous results for $R_\text{final}$ \cite{ma2020role},
it can be verified that for all ER networks or SF networks with $3<\lambda<4$, $\epsilon_I$ has a smaller value in regions with smaller transmissibilities ($T^i$ or $T^b$),
so that the curve of $I_{\max}$~vs.~$r$ goes steeper when $r$ is small.
That is to say,
strategies to reduce $r$ are more effective in controlling peak \jingma{fraction of infected},
if adequate actions are also taken to reduce $T^i$ or $T^b$.
Our results also show that,
compared to the values of $\epsilon_R$ as in $R_\text{final}\propto r^{1/\epsilon_R}$ in each regime \cite{ma2020role},
$\epsilon_I$ is either smaller than or equal to $\epsilon_R$ for all ER networks or SF networks with $3<\lambda<4$.
In general,
as $r$ decreases,
$I_{\max}$ decays faster than $R_\text{final}$ does,
i.e.,
the peak \jingma{fraction of infected} $I_{\max}$ responds more sensitively to $r$ than \jingma{the total fraction of individuals ever been infected $R_\text{final}$}.
Thus,
strategies that reduce $r$ should be prioritized if controlling the peak fraction of infected is more crucial (for example, when medical resources are limited).


\section{Crossovers for $T^i\lesssim 1/(\kappa^i-1)$ when $T^b=1/(\kappa^b-1)$}

Besides the asymptotic behaviors of $I_{\max}$ as $r\rightarrow 0$,
\jingma{we also expect crossovers for $I_{\max}$,
i.e., the relation between $I_{\max}$ and $r$ follows one power law when $r$ is small enough,
but follows a different power law when $r$ is larger.
This happens when the transmissibilities are near the boundaries of different regimes,
for example,
when $T^i\lesssim 1/(\kappa^i-1)$,
i.e., when $T^i$ is smaller than but close to $1/(\kappa^i-1)$.}
Since Regime I is not an epidemic phase,
and the values of $\epsilon_I$ in Regime III and VI are the same,
we will only look at the case when $T^b=1/(\kappa^b-1)$.

When $T^i\lesssim 1/(\kappa^i-1)$,
the relation between $I_{\max}$ and $r$ behaves differently on both sides of the crossover for two reasons.
Firstly,
the behavior of $I_{\max}$ depends on the behavior of $R^b_\text{final}$,
\jingma{and there is a crossover for $R^b_\text{final}$ \cite{ma2020role}.}
This crossover $r_1^*$ is determined by $1/s_{\max} \sim r_1^*(R^b_\text{final})^*$,
\jingma{where $r_1^*$ and $(R^b_\text{final})^*$ are the values of $r$ and $R^b_\text{final}$ when the crossover for $R_\text{final}$ occurs,}
and thus we get the first crossover point $r_1^* \sim |T^i-\frac{1}{\kappa^i-1}|^{[1-(\tau^i-2)(\tau^b-2)]/\sigma}$ when $T^b=1/(\kappa^b-1)$.
Secondly,
whether $I_{\max}$ depends on $R^b_\text{final}$ or $I^b_{\max}$ is determined by how the time scale of a community is compared with that of the whole system.
In this case,
the turning point occurs when the time scale of a community is approximately the same as the time scale of the system,
i.e.,
$|T^i-\frac{1}{\kappa^i-1}|^{-z^i\nu^i} \sim |T^b-T^b_c|^{-z^b\nu^b}$,
which reduces to $|T^i-\frac{1}{\kappa^i-1}| \sim |T^b-T^b_c|$ for ER and SF networks with $3<\lambda<4$,
where $T^b_c$ is the critical value of bridge link transmissibility $T^b$ for the whole system,
given a fixed value of $T^i$.
Then we have the second crossover point (see Appendix \ref{app_sec_scale} for details)
\begin{equation}
\left|T^i-\frac{1}{\kappa^i-1}\right| \sim \frac{\langle k^i \rangle \langle k^b \rangle T^i}{(\kappa^i-1)(\kappa^b-1)^2\big(\frac{1}{\kappa^i-1}-T^i\big)}\cdot r_2^*
,
\end{equation}
which gives $r_2^* \sim |T^i-\frac{1}{\kappa^i-1}|^2$.

\begin{figure}[h!]
\begin{minipage}[t]{0.49\linewidth}\centering
\includegraphics[width=7.5cm]{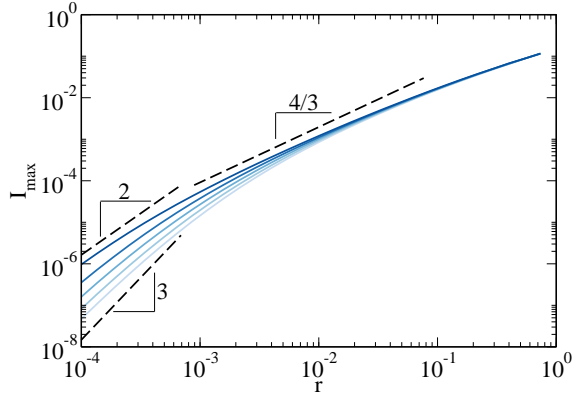}
\medskip
\end{minipage}\hfill
\caption{Crossovers of $I_{\max}$ as a function of $r$ when $T^i\lesssim 1/(\kappa^i-1)$, i.e., $T^i=0.245,0.246,0.247,0.248,0.249$ (from light blue to dark blue), with $T^b=1/(\kappa^b-1)=0.1$. Both internal links and bridge links are ER networks, with $\langle k^i \rangle=4$ and $\langle k^b \rangle=10$, respectively. Dashed lines represent slopes predicted by Tables~\ref{app_tb_regimes_new}~and~\ref{app_tb_regimes_old} for different regimes.}
\label{fig_crossover}
\end{figure}

In summary,
when $T^i\lesssim 1/(\kappa^i-1)$ and $T^b=1/(\kappa^b-1)$,
we expect crossovers in the relation between $I_{\max}$ and $r$.
If $r$ is small enough so that $r<r_1^*$ and $r<r_2^*$,
$I_{\max}$~vs.~$r$ follows its asymptotic behavior as $T^i<1/(\kappa^i-1)$,
while if $r$ is larger than both $r_1^*$ and $r_2^*$,
the relation between $I_{\max}$ and $r$ will be as if $T^i=1/(\kappa^i-1)$,
both of which can be verified by the numerical solutions from Eqs.~(\ref{eq_difference_thetai})-(\ref{eq_difference_deltaib}),
as shown in Fig.~\ref{fig_crossover}.
Moreover,
we can also see a transition part when $r_2^*<r<r_1^*$,
whose slope can also be predicted by combining dependencies of $R_\text{final}$ and $I_{\max}$ with $r$ from different regimes.

Practically, when actions are taken to reduce $r$,
it is essential to know that the peak \jingma{fraction of infected} $I_{\max}$ may not be reduced immediately as fast as the predicted behaviors for asymptotic situations.
This is due to the fact that $r$ may be not small enough,
and we are on the right side of the crossovers.
However,
we would expect $I_{\max}$ to drop down as fast as predicted,
once $r$ goes below the crossover points.
Our results enable us to find the balance between relieving medical pressure and reopening,
and to make better plans for epidemic strategies.


\section{Conclusions}

In this paper,
we study the dynamic properties of a two-community system with bridge nodes,
especially how the peak \jingma{fraction of infected} $I_{\max}$ depends on the fraction of bridge nodes $r$.
We find the asymptotic relation between $I_{\max}$ and $r$ to have power-law behaviors in multiple regimes.
We analytically calculate the power-law exponents for each regime,
which are verified by numerical solutions from the EBCM.
We also find crossovers between regimes when $T^i\lesssim 1/(\kappa^i-1)$ and $T^b=1/(\kappa^b-1)$,
which can be explained by the comparison of time scales between different parts of the system.
Our methodology can be easily extended to situations with multiple communities,
or communities with different internal degree distributions, or different internal transmissibilities.

\jingma{
Our methods can also be adapted to other types of compartmental models,
as long as the final state of the model is R
(i.e., a node is immune to the same disease once recovered),
so that a mapping to the link percolation problem is still available,
such as in the SEIR model,
where E stands for Exposed \cite{gandolfi2013percolation}.
However,
it does not apply to models in which a node may get reinfected,
for example,
the SIS, or SIRS models, etc.
Also note that the mapping from the SIR model to link percolation is exact only when the time to recover after getting infected $t_r$ is fixed,
so that the transmissibilities along links are independent;
otherwise,
the mapping might not be accurate \cite{kenah2007second}.
}

\jingma{
Due to the complexity of the real world,
our model cannot capture all factors in practice,
such as time-varying transmission rate \cite{jagan2020fast,calafiore2020time},
vaccination \cite{alvarez2019dynamic},
or time-dependent epidemic strategies \cite{pham2021estimating,hoertel2021impact},
etc.
Instead,
we focus on one aspect of epidemic strategies,
i.e., the shutting down of international traveling,
and propose new methodology to study and to evaluate epidemic strategies.
Our results serve as an important basis for making epidemic strategies,
e.g.,
to anticipate the effectiveness of a strategy,
and to find the best practice of reopening under the premise that all patients can get timely treatment.
}


\section{Acknowledgments}

J.M. and L.A.B. acknowledge support from DTRA Grant No. 9500309448.
X.M. is supported by the NetSeed: Seedling Research Award of the Network Science Institute of Northeastern University.
L.A.B. wish to thank UNMdP (EXA 956/20) and FONCyT (PICT 1422/2019) for financial support.


\begin{appendices}


\section{EBCM Adapted to the SIR Model}
\label{app_ebcm}

\jingma{The Edge-Based Compartmental Model (EBCM) adapted to the SIR model was first introduced for isolated networks \cite{miller2012edge}
and then extended for multi-community networks with bridge nodes \cite{valdez2018role}.}
In this model,
two auxiliary variables $\theta^i(t),\theta^b(t)$ are defined as the probabilities that the disease has not been transmitted through a randomly chosen internal or bridge link from a node, respectively, by time $t$,
which could fall into one of the three categories:
the node is still susceptible (S) up to this instant (with probability $\Phi_S(t)$),
the node is infected (I) at this instant but has not transmitted through this link yet (with probability $\Phi_I(t)$),
or the node is already recovered (R) and has never transmitted the disease through this link (with probability $\Phi_R(t)$) \cite{valdez2018role}.
\jingma{Recall that $G_{0,1}^{i,b}$ represents generating functions,
where the subscript is used to denote whether the generating function is for the degree distribution ($0$),
or the excess distribution ($1$);
the superscript is $i$ for internal links,
and $b$ for bridge links.}
The time dependence of all variables of the SIR model can then be calculated numerically from \cite{valdez2018role}:

\begin{align}
\theta^i(t+1) {}& = \theta^i(t) - q^i\Phi_I^i(t)
\label{eq_difference_thetai} \\
\theta^b(t+1) {}& = \theta^b(t) - q^b\Phi_I^b(t)
\label{eq_difference_thetab} \\
\begin{split}
\Delta\Phi_S^i(t) {}& = (1-r)\left[G_1^i(\theta^i(t+1)) - G_1^i(\theta^i(t))\right] \\ & \hspace{88pt} + r\left[G_1^i(\theta^i(t+1))G_0^b(\theta^b(t+1)\jingma{)} - G_1^i(\theta^i(t))G_0^b(\theta^b(t))\right]
\end{split}
\label{eq_difference_deltaphisi} \\
\Delta\Phi_S^b(t) {}& = G_0^i(\theta^i(t+1))G_1^b(\theta^b(t+1)\jingma{)} - G_0^i(\theta^i(t))G_1^b(\theta^b(t))
\label{eq_difference_deltaphisb} \\
\Delta\Phi_I^i(t) {}& = - q^i\Phi_I^i(t) - \Delta\Phi_S^i(t) + (1-T^i)\Delta\Phi_S^i(t-t_r)
\label{eq_difference_deltaphiii} \\
\Delta\Phi_I^b(t) {}& = - q^b\Phi_I^b(t) - \Delta\Phi_S^b(t) + (1-T^b)\Delta\Phi_S^b(t-t_r)
\label{eq_difference_deltaphiib} \\
\Delta S^i(t) {}& = (1-r)\left[G_0^i(\theta^i(t+1)) - G_0^i(\theta^i(t))\right]
\label{eq_difference_deltasi} \\
\Delta S^b(t) {}& = r\left[G_0^i(\theta^i(t+1))G_0^b(\theta^b(t+1)) - G_0^i(\theta^i(t))G_0^b(\theta^b(t))\right]
\label{eq_difference_deltasb} \\
\Delta I^i(t) {}& = - \Delta S^i(t) + \Delta S^i(t-t_r)
\label{eq_difference_deltaii} \\
\Delta I^b(t) {}& = - \Delta S^b(t) + \Delta S^b(t-t_r)
\label{eq_difference_deltaib}
\end{align}
where $q^i$ (or $q^b$) is the probability that an infected node transmits the disease to its susceptible neighbor through an internal link (or a bridge link) at each time step,
and $t_r$ is the number of time steps it takes for an infected individual to recover,
and thus $T^i = 1-(1-q^i)^{t_r}$ and $T^b = 1-(1-q^b)^{t_r}$.

Equations~(\ref{eq_difference_thetai})-(\ref{eq_difference_thetab}) are due to the fact that the disease can only \jingma{be} transmitted through a link when the node is infected.
In Eqs.~(\ref{eq_difference_deltaphisi})-(\ref{eq_difference_deltaphisb}) and (\ref{eq_difference_deltasi})-(\ref{eq_difference_deltasb}),
$\Phi_S^i$ or $\Phi_S^b$ is calculated by the probability that the disease has not transmitted to the node through any other links by time $t$,
and $S^i$ or $S^b$ is calculated by the probability that the disease has not transmitted to the node through any of its links by time $t$.
Eqs.~(\ref{eq_difference_deltaphiii})-(\ref{eq_difference_deltaphiib}) and (\ref{eq_difference_deltaii})-(\ref{eq_difference_deltaib})
take $\Delta\theta(t) = \Delta\Phi_S(t)+\Delta\Phi_I(t)+\Delta\Phi_R(t)$ and $0=\Delta S+\Delta I+\Delta R$ into account,
and that all infected nodes will recover after $t_r$ time steps,
so that $\Delta\Phi_R(t)=-\Delta\Phi_S(t-t_r)$ and $\Delta R(t)=-\Delta S(t-t_r)$.


\section{Derivation of $T^b_c$ for the whole system given a fixed $T^i$}
\label{app_sec_scale}

By mapping the final state of the whole system to the giant component in the link percolation process,
we have the self-consistent equations \cite{ma2020role,valdez2018role,son2012percolation}
\begin{eqnarray}
f^i &=& (1-r)\left[1-G_1^i(1-T^if^i)\right] + r\left[1-G_1^i(1-T^i f^i)G_0^b(1-T^bf^b)\right], \\
f^b &=& 1-G_0^i(1-T^if^i)G_1^b(1-T^bf^b), 
\end{eqnarray}
where $f^i$ or $f^b$ is the probability to expand a branch to the infinity through an internal link or a bridge link,
respectively.
\jingma{
The factors $(1-r)$, $r$ and $1$ of each term stand for the fact that the node an internal link leads to has a probability of $(1-r)$ to be an internal node,
and a probability $r$ to be a bridge link,
while bridge links only lead to bridge nodes.}

The critical value of $T^b$ given $T^i$ can be solved by letting the Jacobian matrix satisfy $|J-I|_{f^i,f^b=0}=0$,
where $J_{i,j}=\frac{\partial f_i}{\partial f_j}$,
in which each of $f_i$ and $f_j$ represents $f^i$ or $f^b$.
Thus,
we have
\begin{equation}
    \begin{vmatrix}
      T^i(\kappa^i-1)-1 & rT^b_c\langle k^b \rangle \\[1ex]
      T^i\langle k^i \rangle & T^b_c(\kappa^b-1)-1 \\[1ex]
    \end{vmatrix}
    = 0.
\end{equation}
So $T^{b}_{c}$ is given by
\begin{equation}
\label{eq_tbc}
T^b_c = \frac{(\kappa^i-1)T^i-1}{(\kappa^b-1)\big[(\kappa^i-1)T^i-1\big]-r\langle k^i \rangle \langle k^b \rangle T^i}.
\end{equation}

When $r$ is small,
this expression can be approximated by looking only at the first two orders of its Taylor series expansion around $r=0$.
The zeroth order gives
\begin{equation}
\left.T^b_c\right|_{r=0} = \frac{1}{\kappa^b-1},
\end{equation}
while the first order derivative is
\begin{equation}
\label{app_eq_rbbelow}
\begin{split}
\left.\frac{\partial T^b_c}{\partial r}\right|_{r=0}
& = \left.\frac{\big[(\kappa^i-1)T^i-1\big]\langle k^i \rangle \langle k^b \rangle T^i}{\Big[(\kappa^b-1)\big[(\kappa^i-1)T^i-1\big]-r\langle k^i \rangle \langle k^b \rangle T^i\Big]^2}\right|_{r=0} \\
& = \frac{\langle k^i \rangle \langle k^b \rangle T^i}{(\kappa^b-1)^2\big[(\kappa^i-1)T^i-1\big]} \\
& = -\frac{\langle k^i \rangle \langle k^b \rangle T^i}{(\kappa^i-1)(\kappa^b-1)^2\big[\frac{1}{\kappa^i-1}-T^i\big]}.
\end{split}
\end{equation}
Thus, we have
\begin{equation}
T^b_c \approx \frac{1}{\kappa^b-1} - \frac{\langle k^i \rangle \langle k^b \rangle T^i}{(\kappa^i-1)(\kappa^b-1)^2\left[\frac{1}{\kappa^i-1}-T^i\right]}\cdot r .
\end{equation}

\jingma{
\section{Phase diagram and regimes}
\label{app_sec_diagram}
}

\begin{figure}[htb]
\centering
\includegraphics[width=0.6\textwidth]{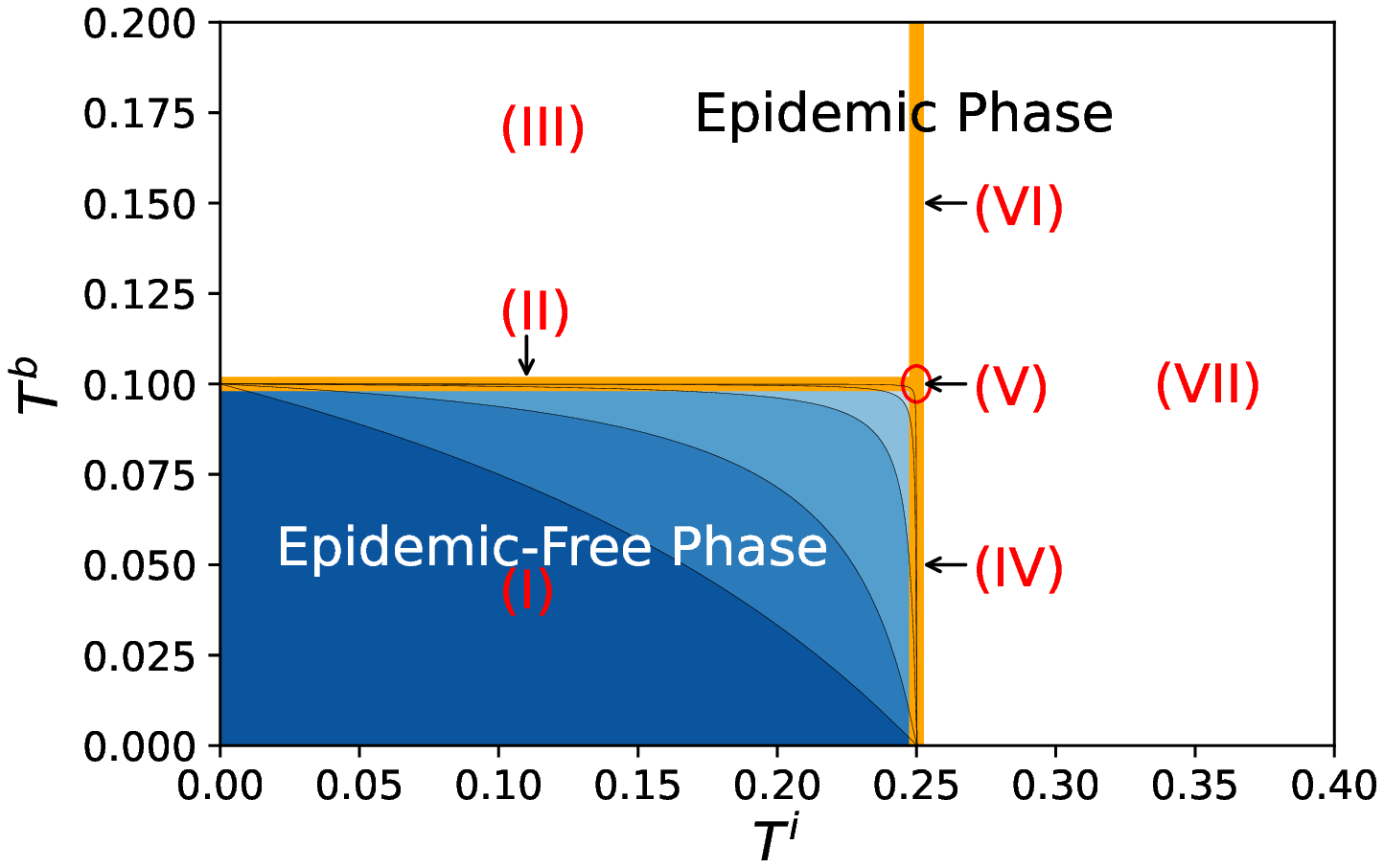}
\caption{Phase diagram for two ER communities connected by ER bridge links, with $\langle k^i \rangle=4$ and $\langle k^b \rangle=10$. The blue shaded areas under the curves of $T^b_c$ are the nonepidemic phases for different values of $r$ ($r=0.5,0.1,0.01,0.001,0.0001$, from dark blue to light blue). As $r\rightarrow 0$, the nonepidemic phase expands and tends to be a rectangle,
whose boundaries are given by
$T^i = 1/(\kappa^i-1)$ and $T^b=1/(\kappa^b-1)$ \cite{ma2020role}.}
\label{fig_regimes}
\end{figure}

\jingma{
The phase diagram (Fig.~\ref{fig_regimes}) of the bridge link transmissibility $T^b_c$ given in Eq.~(\ref{eq_tbc}) was originally presented in our previous work \cite{ma2020role}.
As an example,
for two ER communities connected by ER bridge links with $\langle k^i \rangle=4$ and $\langle k^b \rangle=10$,
we can see that as $r\rightarrow 0$,
the nonepidemic phase (the shaded area below the curve of $T^b_c$) tends to be a rectangle.
The boundaries of the rectangle are given by
$T^i = 1/(\kappa^i-1)$ and $T^b=1/(\kappa^b-1)$,
along which the whole space is split into several regimes \cite{ma2020role}.
}

\section{$I_{\max}$ with $n$ patient zeros in an isolated network}
\label{app_sec_multiple}

In the case where a disease starts spreading from one patient,
i.e., ``patient zero'',
in an isolated network,
there are behaviors around criticality
$\langle s \rangle \sim |T-T_c|^{-\gamma}$,
where $\langle s \rangle$ represents the mean cluster size,
$l \sim |T-T_c|^{-z\nu}$,
where $l$ represents the chemical distance or shortest-path distance,
and thus $\langle s \rangle\sim l^{\gamma/z\nu}$.
Considering $\langle s \rangle \sim \int N(l)dl$ and $I\sim N(l)$,
we have $I \sim l^{\gamma/z\nu-1}$ \cite{zhou2012shortest},
which becomes $I \sim O(1)$
for both ER and SF networks with $3<\lambda<4$,
whose $\gamma=1$, $z=2$ and $\nu=1/2$
\cite{cohen2003structural}.

For epidemics starting from $n$ patient zeros simultaneously in an isolated network,
$I$ would be less than $n\cdot O(1)$ with time going on, if the spreading paths from different patient zeros overlap.
Due to the initial condition $I_0=n$,
we will have $I_{\max} \propto n$.

\end{appendices}

\bibliography{main}

\end{document}